\documentstyle[12pt]{article}
\def\slash#1{\setbox0=\hbox{$#1$}#1\hskip-\wd0\hbox to\wd0{\hss\sl/\/\hss}}
\begin{document}
\baselineskip=20 pt
\def\l{\lambda}
\def\L{\Lambda}
\def\b{\beta}
\def\mphi{m_{\phi}}
\def\hphi{\hat{\phi}}
\def\vphi{\langle \phi \rangle}
\def\etamunu{\eta^{\mu\nu}}
\def\dmul{\partial_{\mu}}
\def\dnul{\partial_{\nu}}
\def\bea{\begin{eqnarray}}
\def\eea{\end{eqnarray}}
\def\bfl{\begin{flushleft}}
\def\efl{\end{flushleft}}
\def\bgt{\beta (g_t)}
\def\bgtmz{\beta (g_t(m_z))}
\def\blam{\beta (\lambda )}
\def\vphic{\vphi_{crit}}
\def\mphi{m_{\phi}}
\begin{center} 

{\large \bf
  Constraints on radion vev from the beta functions \\
 
\vskip 0.04in

 for $g_t$ and $\l$ in the Randall-Sundrum model }

\end{center}

\vskip 10pT   
\begin{center}
{\large\sl \bf{Prasanta Das}~\footnote{E-mail: pdas@iitk.ac.in}
}
\vskip  5pT  
{\rm
Department of Physics, Indian Institute of Technology, \\
Kanpur 208 016, India.} \\
\end{center}


\begin{center}
{\large\sl  \bf{Uma Mahanta}~\footnote{E-mail:mahanta@mri.ernet.in}
}
\vskip 5pT
{\rm
Mehta Research Institute, \\
Chhatnag Road, Jhusi
Allahabad-211019, India .}\\
\end{center}

\centerline{\bf Abstarct}

In this paper we determine how the beta  function for the top Yukawa
coupling to one loop is modified by a light stabilized radion in the
Randall-Sundrum model.
We then use this beta function together with $\beta(\l)$ to determine 
lower bounds on the radion vev $\vphi$, for different values of $\l(m_z)$
and $\mphi$
by demanding that both $g_t(\mu)$ and $\l(\mu)$ should remain perturbative
from $m_z$ up to the cut off $\L$. We find that the lower bounds on
$\vphi$ that follow from the condition $\l(\L) < \sqrt{4 \pi}$ nearly
agree with those that follow from $g_t (\L) < \sqrt{4 \pi}$. We have also
determined the critical value of $\vphi$ for different values of $\l(m_z)$
and $m_{\phi}$ from the line of fixed points which is given by
$\beta(g_t) = 0$

\newpage

\bfl
{\large {\bf {Introduction}}}
\efl

Recently several attractive proposals based on theories of extra
dimensions \cite{ADD} have been put forward to explain the hierarchy
problem of
the Standard Model(SM) to explain this hierarchy problem. Among them, the
Randall-Sundrum model \cite{RS} is particularly interesting because it
considers a five dimensional world based on the following non-factorizable
metric

\bea
ds^2= e^{-2k r_{c} |\theta | }{\eta_{\mu\nu}} dx^{\mu} dx^{\nu}-
{r_c}^2 d{\theta}^2
\eea

Here $r_c$ measures the size of the extra dimension which is an ${S^1/Z_2}$
orbifold. $x^{\mu }$ are the coordinates of the four
dimensional space-time. $-\pi\le \theta \le \pi$ is the coordinate of the
extra dimension
with $\theta$ and $-\theta$ identified. k is a mass parameter of the order
of the fundamental five dimensional Planck mass M. 
Two 3 branes are placed at the orbifold
fixed points $\theta =0$ (hidden brane) and $\theta =\pi$ (visible
brane). Randall and Sundrum showed that any field on the visible brane
with a fundamental 
 mass parameter $m_0$ gets an effective mass $$m = m_0 e^{-k r_c \pi}$$
due to the exponential warp factor. Therefore
 for $k r_c \approx 14$ the electroweak scale $\L_{ew}$
is generated from the Planck scale $M_{pl}$ by the warp factor.

In the Randall-Sundrum model $r_{c}$ is the vacuum expectation 
value (vev) of a 
massless scalar field T(x). The modulus was therefore not stabilized
by some dynamics. In order to stabilize the modulus Goldberger and Wise
\cite{GW} introduced a scalar field $\chi(x, \theta )$ in the bulk 
with interaction
potentials localised on the branes. This they showed could generate a
potential for $T(x)$ and stabilize the modulus at the right value 
($k r_{c} \approx 14$) needed to explain 
 the hierarchy between $M_{pl}$  and  $\L_{ew}$ without any excessive
fine tuning of the parameters of the model.

In the Randall-Sundrum model the SM fields are assumed to be localized
on the visible brane at $\theta =\pi$. However the SM action is modified
due to the exponential warp factor. Small fluctuations of the modulus
field from its vev gives rise to non-trivial couplings  of 
the modulus field with the SM fields. In this paper we shall derive
the couplings of a stabilized radion to the top quark lagrangian upto
quadratic order in $\frac{\hat{\phi}}{\vphi}$.Here $\hphi$ is a small
fluctuation of the 
radion field from its vev and is given by $\phi =f e^{-k\pi T(x)}=
\vphi +\hphi$. $\vphi =f e^{-k\pi r_c}$ is the vev of $\phi$ and f
is a mass parameter of the order of M. We then use the radion couplings to
the top quark to determine the modification in the beta function for $g_t$
to one loop due to a light stabilized radion. 
Radion phenomenology in the Randall-Sundrum model depends on two 
unknown parameters, the radion mass $\mphi$ and its vev $\vphi$. 
Determining bounds on these two parameters is therefore extremely
important. Consistency with the collider data
\cite{MR} requires that $\vphi$ must
be of the order $v$(higgs vev) or greater. In contrast the radion mass
remains relatively unconstrained by the low energy phenomenology.
 In this paper we shall use the beta
functions for $g_t$ and $\l$ in the presence of alight stabilized radion
 to derive lower bounds on $\vphi$ for
different values of $\l(m_z)$ and $m_\phi$ by requiring that the running
couplings $g_t(\mu)$ and $\l(\mu)$ should remain perturbative from $m_z$
all the way
up to the cut off $\L$. In particular we have determined the lower bound
on $\vphi$ for the case $m_h(m_z) = 114~ GeV$ which corresponds to the
present direct bound on $m_h$. We find that the lower bounds on $\vphi$
that follow from the condition $g_t (\L ) < \sqrt{4 \pi}$ nearly agree withn
those that follow from $\l(\L) < \sqrt{4 \pi}$. Finally we have drawn the
line of fixed points, which is obtained by setting $\beta(g_t) = 0$, in
the $\vphi$ vs $m_h (m_z)$ plane for different fixed values of $m_\phi$. 
We show that these curves could also be used to determine lower(upper)
bounds on $\vphi$ depending upon whether $\beta(g_t(m_z))$ is negative
(positive). For the sake of simplicity, in this paper we shall not
consider the phenomenological
 effects of any curvature scalar-higgs mixing operator.

\vspace*{0.25in}

\bfl
{\large {\bf {Radion couplings to the top quark in the Randall-Sundrum
model}}}
\efl

 The couplings of the radion to the top quark in the Randall-Sundrum model
can be derived from the following action

\bea
S_1 = \int d^4 x \sqrt{- g_v} \left[{\overline{\psi}}\left(i
\gamma_a e^{a\mu}D_\mu - m \right)\psi - \frac{g_t}{\sqrt{2}} H
{\overline{\psi}} \psi \right] 
\eea

where $e^{a\mu}$ is the contravariant vierbein field for the visible
brane. In the presence of radion fluctuation it satifies the normalization
condition 
\bea
e^{a\mu}e_{a}^{\nu} = g^{\mu\nu} =
\left(\frac{\phi}{f}\right)^{-2}~\eta^{\mu\nu} =
e^{2 \pi k T(x)}~\eta^{\mu\nu}
\eea
$D_\mu$ is the covariant derivative with respect to general coordinate
transformation and is given by 
$$D_\mu \psi = \partial_\mu \psi + \frac{1}{2}w_{\mu}^{ab} \Sigma_{ab}
\psi$$ 
$w_{\mu}^{ab}$ being the spin connection. It can be computed 
from the vierbein fields by using the relation \cite{MV},

\bea
w_{\mu}^{ab} = \frac{1}{2} e^{\nu a}(\partial_\mu e^b_\nu - \partial_\nu
e^b_\mu) - \frac{1}{2} e^{\nu b}(\partial_\mu e^a_\nu -
\partial_\nu e^a_\mu) -  \frac{1}{2} e^{\rho a}e^{\sigma b}(\partial_\rho
e_{\sigma c} - \partial_\sigma e_{\rho c})e^c_\mu 
\eea

$\Sigma_{ab}$ is given by the expression $\Sigma_{ab} = \frac{1}{4}
\left[\gamma_a,\gamma_b \right]$. It can be shown that in the presence of
radion fluctuations on the visible brane, the spin connection is given by,
\bea
w_{\mu}^{ab} = \frac{1}{\phi} \partial_\nu \phi \left[e^{\nu
b}e^a_\mu - e^{\nu a} e^b_\mu\right]
\eea
The covariant derivative of the fermion field then becomes 
$$
D_\mu \psi = \partial_\mu \psi + \frac{1}{4
\phi}\partial^\nu \phi \left[\gamma_\mu,\gamma_\nu\right] \psi
$$
where the $\gamma_\mu$ are independent of space time coordinates. The
action that determines the radion couplings to top quark can therefore be
written as
\bea
S_1 = \int d^4 x \left(\frac{\phi}{f}\right)^4
\left[\left(\frac{\phi}{f}\right)^{-1} {\overline{\psi}}\{i
\gamma^\mu \partial_\mu + \frac{3 i}{2 \phi} \partial_\mu \phi
\gamma^\mu \}\psi - m_t {\overline{\psi}} \psi -
\frac{g_t}{\sqrt{2}} H {\overline{\psi}} \psi \right] \noindent
\nonumber \\
= \int d^4 x \left[{\overline{\tilde{\psi}}}\{i \gamma^\mu \partial_\mu
{\tilde \psi} + \frac{3 i}{2 \phi}\partial_\mu \phi \gamma^\mu
{\tilde \psi} \}\left(1 + \frac{\hat{\phi}}{\vphi}\right)^3 -
\left(\tilde{m_t} + \frac{g_t}{\sqrt{2}} \tilde{H}\right)\left(1 +
\frac{\hat{\phi}}{\vphi}\right)^4 {\overline{\tilde{\psi}}} {\tilde \psi}
\right] \noindent
\nonumber \\
= \int d^4 x \left[{\overline{\tilde{\psi}}}i \gamma^\mu \partial_\mu
{\tilde \psi} - \tilde{m_t} {\overline{\tilde{\psi}}} {\tilde{\psi}} -
\frac{g_t}{\sqrt{2}} \tilde{H} {\overline{\tilde{\psi}}} {\tilde
\psi}\right] \noindent
\nonumber \\
+  \int d^4 x \left[ \frac{3 i}{ \vphi} {\overline{\tilde{\psi}}}
\gamma^\mu \partial_\mu {\tilde \psi}~ {\hat{\phi}} + \frac{3 i}{2 \vphi}
{\overline{\tilde{\psi}}} \gamma^\mu {\tilde \psi}~ \partial_\mu
{\hat{\phi}} - 4 \left(\tilde{m_t} + \frac{g_t}{\sqrt{2}} \tilde{H} 
\right)\frac{\hat{\phi}}{\vphi}{\overline{\tilde{\psi}}} {\tilde
\psi}\right] \noindent
\nonumber \\
+  \int d^4 x \left[ 3~ {\overline{\tilde{\psi}}}i
\gamma^\mu \partial_\mu {\tilde \psi}~ \frac{\hat{\phi}^2}{\vphi^2} + 
\frac{3 i}{\vphi^2}~ {\hat{\phi}}~ {\overline{\tilde{\psi}}}\gamma^\mu
{\tilde \psi}~ \partial_\mu {\hat{\phi}} - 6 \left(\tilde{m_t} 
+ \frac{g_t}{\sqrt{2}} \tilde{H}
\right)\frac{\hat{\phi}^2}{\vphi^2}{\overline{\tilde{\psi}}}
{\tilde \psi}\right]
\eea
Here,  $\psi = \left(\frac{f}{\vphi}\right)^{3/2} {\tilde{\psi}}$, 
~$H = \left(\frac{f}{\vphi}\right) {\tilde{H}}$~ and~ 
$m = \left(\frac{f}{\vphi}\right) {\tilde{m}}$.

In the following we shall assume that all fields and parameters have been
properly scaled so as to corresponds to the TeV scale and drop the
$\it{tilde}$ sign.



In order to determine the radion contribution to $\beta(g_t)$ we also need
the radion couplings to the higgs scalar which can be determined from the
following action

\bea
S_2 = \int d^4 x \left[\frac{1}{2} \partial_\mu h  \partial^\mu h~\left(1
+ \frac{\hat{\phi}}{\vphi}\right)^2  - \frac{\l}{4}\left(4
h^2 v^2 + 4 h^3 v + h^4 \right) \left(1 +
\frac{\hat{\phi}}{\vphi}\right)^4 \right] 
\eea

The radion couplings to the K.E. of the higgs boson will contribute only
to $Z_h$ (wave function renormalization constant of higgs boson) but not
to the renormalization of $H {\overline{\psi}} \psi$ vertex.



\bfl
{\large {\bf {Radion contribution to the beta function for $g_t$}}}
\efl
 Radion contribution to $\beta(g_t)$ arises from two different 
sources:(a) renormalization of $H {\overline{\psi}} \psi$ vertex due to
radion and (b) wave function renormalization constants of top quark and
higgs boson due to radion.

 Radion contribution to the $H {\overline{\psi}} \psi$ vertex correction
arises from the Feynman diagrams shown in Fig. 2.


 In order to determine the contribution of these diagrams to
$H{\overline{\psi}} \psi$ vertex correction we have to consider only those
terms in the loop integral that do not depend on the external momentum.
The reason being external momentum will give rise to derivative of
external fields and there are no such derivatives in the Yukawa term
$H{\overline{\psi}} \psi$. Considering only the external momentum
independent terms and retaining only the contributions of 
such terms thatdiverge with
the cut off $\L$ we get 

\bea
\Gamma_1 = -\left(\frac{g_t}{\sqrt{2}}\right) \frac{1}{16 \pi^2
\vphi^2}\left[\frac{9}{4} \L^2 - \frac{1}{4}(9 m_\phi^2 - 5
m_t^2)~ln{\frac{\L^2}{\mu^2}}\right] 
\eea
\bea
\Gamma_2 = \left(\frac{g_t}{\sqrt{2}}\right) \frac{1}{16 \pi^2
\vphi^2}\left[16~ m_h^2~ ln{\frac{\L^2}{\mu^2}}\right]
\eea
\bea
\Gamma_3 = -6 \left(\frac{g_t}{\sqrt{2}}\right) \frac{1}{16 \pi^2 
\vphi^2}\left[\L^2 -  m_\phi^2 ~ln{\frac{\L^2}{\mu^2}}\right] 
\eea
\bea
\Gamma_4 = - 12 \left(\frac{g_t}{\sqrt{2}}\right) \frac{1}{16 \pi^2   
\vphi^2}\left[m_h^2~ ln{\frac{\L^2}{\mu^2}}\right]
\eea
\bea
\Gamma_5 = - \left(\frac{g_t}{\sqrt{2}}\right) \frac{1}{16 \pi^2
\vphi^2}\left[- 12 \L^2 + (12 m_\phi^2 - 20 m_t^2) 
~ln{\frac{\L^2}{\mu^2}}\right]
\eea
Here $\mu$ is the renormalization  mass-scale. 
 
 The wave function renormalization constants $Z_h$ and $Z_t$ of the higgs
boson and top quark arise from the Feynman diagrams shown in Fig 3.

By considering the terms proportional to $p^2$ of Fig 3a and the terms 
proportional to $\slash{p}$ of Fig 3b it can be shown that

\bea
Z_h =  1 +  \frac{1}{32 \pi^2 \vphi^2}\left[7 m_h^2 + m_\phi^2 \right]
~ln{\frac{\L^2}{\mu^2}}
\eea
and
\bea
Z_\phi =  1 +  \frac{1}{16 \pi^2 \vphi^2}\left[ \frac{39}{8} \L^2 - 
6 m_\phi^2 ~ln{\frac{\L^2}{\mu^2}} + \frac{13}{4} m_t^2
~ln{\frac{\L^2}{\mu^2}} \right]      
\eea

Using the vertex and wave function renormalization constants given above
it can be shown that the radion contribution $g_t^r(\mu )$
 to the renormalized Yukawa
coupling is given by,
\bea
g_t^r(\mu) =  \frac{g_t}{16 \pi^2 \vphi^2}\left[ \frac{9}{8} \L^2 - 2
m_\phi^2~ln{\frac{\L^2}{\mu^2}} - \frac{31}{2}
m_t^2~ln{\frac{\L^2}{\mu^2}} -  \frac{9}{4}m_h^2~ln{\frac{\L^2}{\mu^2}} 
\right]
\eea

The complete beta function for $g_t$ in the presence of radion
fluctuations then becomes
\bea
\beta(g_t(\mu)) = \beta_{SM}(g_t(\mu))+  
\frac{g_t}{16 \pi^2 \vphi^2}\left[ 4 m_\phi^2 + \frac{31}{2} g_t^2 v^2 + 9
\l v^2 \right]
\eea                           
where \cite{BHL}

 $$\beta_{SM}(g_t(\mu)) = \frac{g_t}{16 \pi^2}
\left[\frac{9}{2} g_t^2 - 8 g_3^2 - \frac{9}{4} g_2^2 -
\frac{17}{12} g_1^2\right]$$

We would like to note that the radion contribution to $\beta(g_t)$ is
positive definite. On the contrary in the context of the SM, $\beta(g_t)$
is negative from $m_z$ all the way up to a cut off $\L$ in the few tens of
TeV. However since the radion contribution to $\beta(g_t)$ is positive,
in the presence of a light radion $\beta(g_t)$ 
 can be positive or negative depending on the values of $m_\phi$,
$m_h$ and $\vphi$.
Note that $\beta(g_t)$
besides depending on $g_t$ also depends on $\l$ and the three SM gauge
couplings $g_1$, $g_2$ and $g_3$. We need to know the beta functions of
the later couplings also in order to solve for $g_t(\mu)$. It was shown in
Ref \cite{DM}  that beta function for $\l$ in the presence of a stabilized
radion to one loop is given by
\bea
{\beta (\lambda )}  = \mu {d\lambda \over d\mu }={1\over
8\pi^2}[9\lambda^2 +
{402 \lambda^2 v^2\over \vphi^2}+ {144\lambda^2 v^4\over \vphi^4} +
{5\lambda \mphi^2 \over \vphi^2}
+\lambda (6 g_y^2-{9\over 2} g_2^2-{3\over 2}g_1^ 2)]
\nonumber \\
+\frac{1}{8 \pi^2} [-6 g_y^4+{3\over 16} (g_2^4+{1\over 2}(g_2^2+g_1^
2)^2)]
\eea

The beta functions for $g_1$, $g_2$ and $g_3$ to one loop are given by the
same expressions as in the SM. In this paper we shall assume for
simplicity that the radion mass $m_\phi$ does not run with $\mu$.

\bfl
{\large {\bf {Lower bound on $\vphi$ from the condition 
$g_t (\L)<\sqrt{4 \pi}$}}}
\efl

 The radion contribution to $\beta(g_t)$ being positive and inversely
proportional to $\vphi^2$. It is possible that for small enough $\vphi$
the running coupling $g_t(\mu)$ by starting from its known value
$g_t(m_z)$ could exceed the perturbative bound $g_t (\L)<\sqrt{4 \pi}$ at
the cut off $\L$. This characteristic of the RG evolution of $g_t(\mu)$ in
the presence of a light radion can be used to determine lower bounds on
$\vphi$ for different fixed values of $\l(m_z)$ and $m_\phi$. Throughout
this paper we shall use the following initial condition on $g_t$, $g_1$,
$g_2$ and $g_3$ : $g_t(m_z) = 1.001$, $g_1(m_z) = 0.356$, $g_2(m_z) =
0.644$ and $g_3(m_z) = 1.218$. Since the higgs and radion masses are as
yet unknown we shall consider different values of $\l(m_z)$ and $m_\phi$
in our analysis. Using the beta functions for $g_t$, $\l$, $g_1$, $g_2$
and $g_3$ together with the above initial conditions one can determine the
values of $g_t(\mu)$ for any value of $\mu$ between $m_z$ and the cut off
$\L$. Here we shall assume that the cut off is related to the radion vev
$\vphi$ by the usual naive dimensional analysis estimate $\L = 4 \pi
\vphi$. In Fig.4a and 4b we have plotted the values of $g_t(\L)$ against
the radion vev $\vphi$ for different values of $m_h(m_z)$ and $m_\phi$. 

We find that as long as $\vphi$ remains greater than some lower bound,
$g_t(\L)$ is practically independent of $\vphi$ and the curve is almost
horizontal. However as the lower bound is approached $g_t(\L)$ increases
very sharply and the curve becomes almost vertical. In fact $g_t(\L)$ hits
the Landau pole at the lower bound of $\vphi$. Note that different points
on the vertical part of the curve correspond to different ultraviolet
boundary conditions on $g_t$ but to the same low energy values of
$m_\phi$, $m_h$ and $\vphi$.
 Table.1 gives the values of the lower bound on
$\vphi$ for different values of $m_\phi$ and $m_h(m_z)$ obtained from the
condition $g_t(\L) < \sqrt{4 \pi}$.

\vspace*{-0.20in}

\begin{center}
Table.1
\end{center}
\vspace*{-0.25in}

\begin{center}
\begin{tabular} {|c|c|c|c|c|}\hline \hline
\multicolumn{1}{|c}{\bf $m_\phi(GeV)$} &
\multicolumn{4}{|c|}{\bf $m_h(m_z)$ ($GeV$)} \\ \cline{2-5}
 & 150 & 200 & 250 & 300 \\ \hline \hline
 50 & 405 & 665 & 981 & 1428 \\ \hline \hline
 300 & 413 & 675 & 991 & 1438 \\ \hline \hline
\end{tabular} 
\end{center}

\vspace*{0.2in}  

 We find from Table.1 that the lower bound on $\vphi$ increases
with $m_h(m_z)$  but is practically independent of $m_\phi$ as long as the
radion is light. The reason being the term involving $m_h^2$ in 
$\delta\beta(g_t)$ has a larger positive coefficient and is inversely
proportional to $\vphi^2$. 

\newpage 

\bfl
{\large {\bf {Lower bound on $\vphi$ from the condition
$\l (\L)<\sqrt{4 \pi}$}}}
\efl

Lower bounds on $\vphi$ can also be determined from the condition that
$\l(\L)$ must be less than the perturbative limit $\sqrt{4 \pi}$. In
Fig.5(a) and 5(b) we have plotted $\l(\L)$ against $\vphi$ for different
values of $m_\phi$ and $m_h(m_z)$. The $\l(\L)$ vs $\vphi$ curves look
very similar to the $g_t(\L)$ vs $\vphi$ curves. As long as $\vphi$ is
greater than some lower bound, $\l(\L)$ is practically independent of
$\vphi$. However as the lower bound is approached $\l(\L)$ is increases
very sharply and the curves becomes almost vertical. We also find that the
rise becomes sharper and sharper with decreasing $m_h(m_z)$. Of
particular importance is the curve corresponding to $m_h(m_z) = 114~ GeV$ 
which corresponds to the present lower bound on $m_h$ from LEPII \cite{TJ}
This curve is almost vertical. Different points on the curve corresponds
to different ultraviolet boundary conditions (UVBC) on $\l(\L)$ but to the
same low energy values of $m_\phi$, $m_h(\mu)$ and $\vphi$. The fact that
the low energy values of $m_\phi$, $m_h(\mu)$ and $\vphi$ are insensitive
to the UVBC on $\l(\L)$ implies that some kind of infrared fixed point
scnerio is in operation here. The lower bound on $\vphi$ corresponding to
$m_\phi(m_z) = 114~GeV$ is about $243~GeV$. Table.2 gives the lower bound
on $\vphi$ obtained from the condition $\l(\L) < \sqrt{4 \pi}$
for different values of $m_\phi$ and $m_h(m_z)$.

\vspace*{-0.20in}

\begin{center}
Table.2
\end{center}
\vspace*{-0.25in}

\begin{center}
\begin{tabular} {|c|c|c|c|c|c|}\hline \hline
\multicolumn{1}{|c}{\bf $m_\phi(GeV)$} &
\multicolumn{5}{|c|}{\bf $m_h(m_z)$ ($GeV$)} \\ \cline{2-6}
 & 114 & 150 & 200 & 250 & 300 \\ \hline \hline
 50 & 242 & 408 & 689 & 1078 & 1816 \\ \hline \hline
 300 & 246 & 416 & 699 & 1090 & 1832 \\ \hline \hline
\end{tabular} 
\end{center}

\newpage

\bfl
{\bf Bounds on $\vphi$ from the sign of $\bgt$}
\efl
In the purely SM $\bgtmz$ is negative. However in the presence of a  
light stabilized radion, $\bgtmz$ can be positive or negative depending
on whether $\vphi$ is smaller than or greater than some critical value
$\vphic$. 
The critical value of $\vphic$ or more generally the critical
line can be determined from the equation $\bgtmz =0$. In Fig 6 we have
plotted the values of $\vphic$ obtained by solving the equation
$\bgtmz =0$ against $m_h(m_z)$ for three different values of $\mphi$.
The region above any particular curve corresponds to $\bgtmz <0$ and
the region below the curve corresponds to $\bgtmz >0$. The purely SM
($\vphi \rightarrow \infty$)
corresponds to a region that lies very far above 
$\vphi \rightarrow \infty$ the curve. Fig.6 shows that the value 
of $\vphic$ increases both with increasing $m_h(m_z)$ and $\mphi$.

\newpage

\bfl
{\bf Acknowledgement}
\efl
We would like to thank Dr. B. Ananthanarayan for helping us with his
code for solving the RG eqns for $g_t$ and $\lambda$.
Uma Mahanta would like to thank the Department of Physics of IIT
Kanpur for hospitality and support while this work was in progress.

\newpage


\begin{figure}[htb]
\begin{center}
\vspace*{1.7in}
      \relax\noindent\hskip -5.4in\relax{\includegraphics{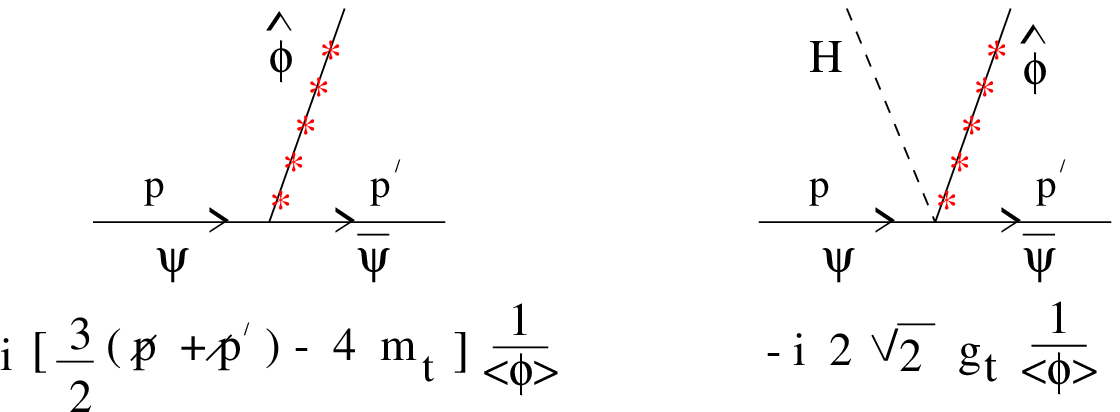}}
\end{center}
\end{figure}
\vspace*{-0.2in}
\noindent { Figure. 1a}.
{\it { Feynman rules for one radion coupling to top quark.}}


\begin{figure}[htb]
\begin{center}
\vspace*{1.9in}
      \relax\noindent\hskip -5.4in\relax{\includegraphics{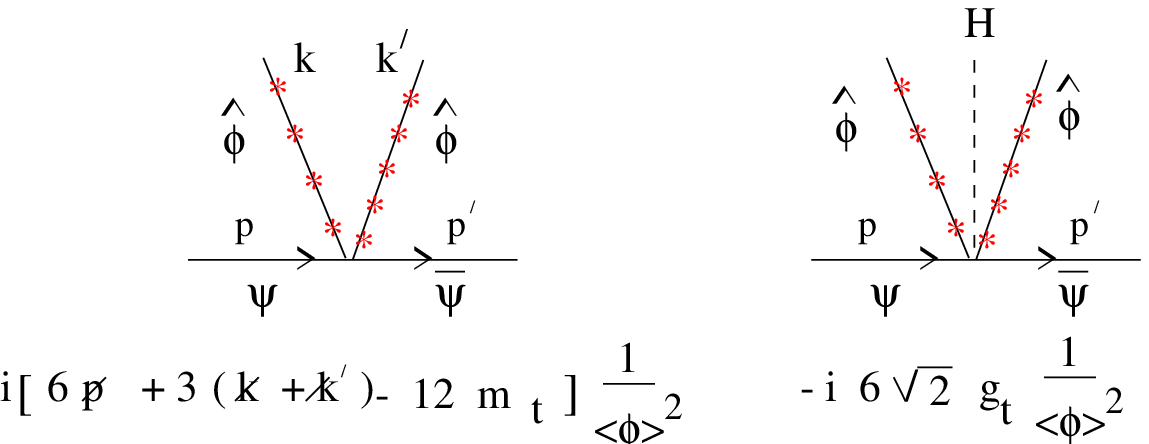}}
\end{center}
\end{figure}
\vspace*{-0.2in}
\noindent { Figure. 1b}.
{\it { Feynman rules for two radion coupling to top quark.}}

\newpage

\begin{figure}[htb]
\begin{center}
\vspace*{2.0in}
      \relax\noindent\hskip -5.4in\relax{\includegraphics{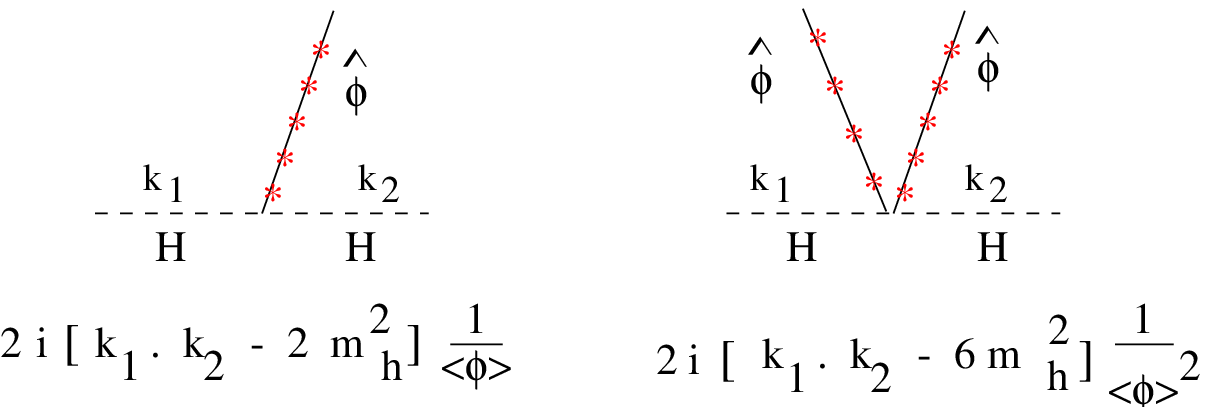}}
\end{center}
\end{figure}
\vspace*{-0.2in}
\noindent { Figure. 1c}.
{\it { Feynman rules for one and two radion couplings to higgs boson.}}

\newpage

\vspace*{0.25in}

\begin{figure}[htb]
\begin{center}
\vspace*{4.5in}
      \relax\noindent\hskip -6.4in\relax{\includegraphics{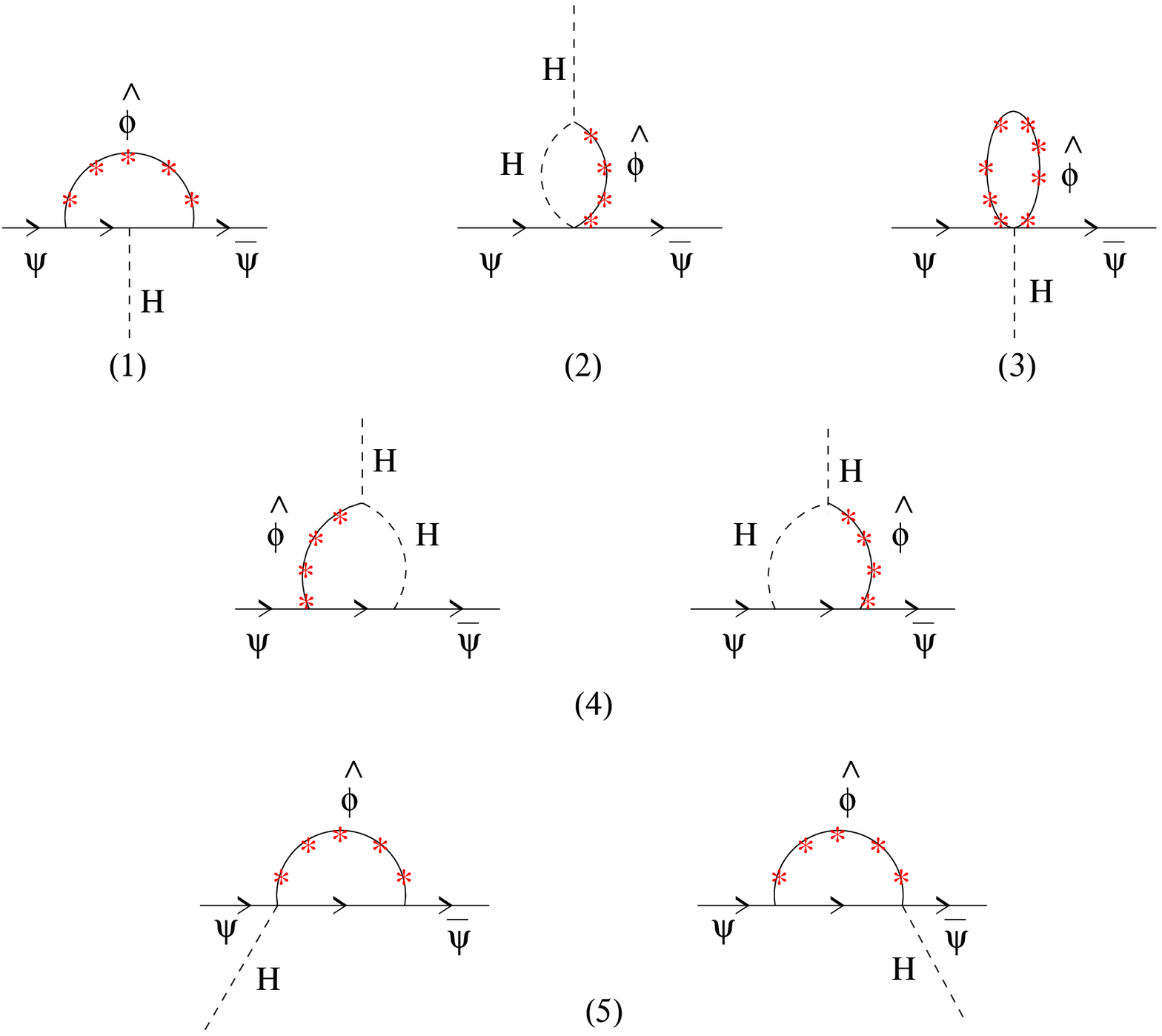}} 
\end{center}
\end{figure}   
\vspace*{-0.2in}
\noindent { Figure. 2}.
{\it { Feynman diagrams that give rise to radion contribution to
$H{\overline{\psi}} \psi$ vertex correction.}}

\newpage

\vspace*{-0.5in}

\begin{figure}[htb]
\begin{center}
\vspace*{2.0in}
      \relax\noindent\hskip -4.4in\relax{\includegraphics{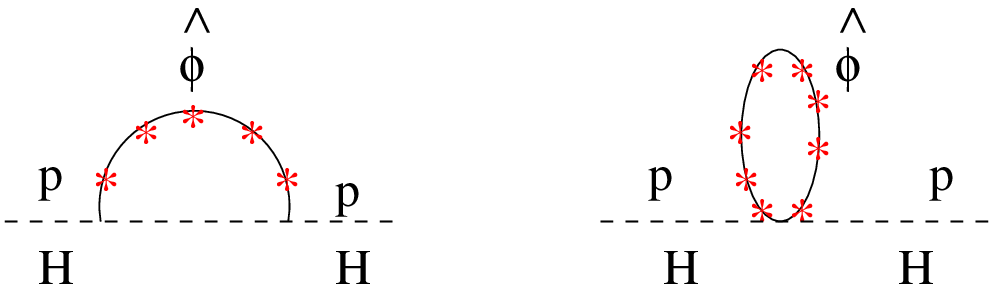}}
\end{center}
\end{figure}
\vspace*{-0.2in}
\noindent { Figure. 3a}.
{\it { Feynman diagrams giving rise to $Z_h$.}}

\vspace*{0.25in}

\begin{figure}[htb]
\begin{center}
\vspace*{2.0in}  
      \relax\noindent\hskip -4.4in\relax{\includegraphics{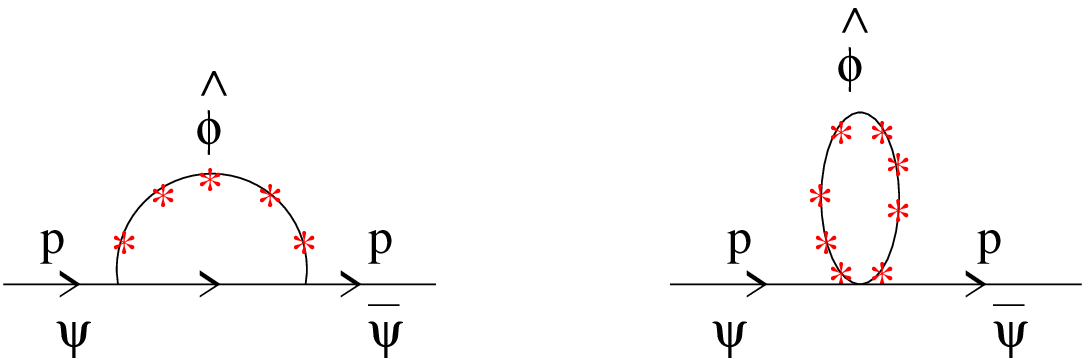}}
\end{center}
\end{figure}
\vspace*{-0.2in}
\noindent { Figure. 3b}.
{\it { Feynman diagrams giving rise to $Z_t$.}}

\newpage

\begin{figure}[htb]
\begin{center}
\vspace*{4.5in}
      \relax\noindent\hskip -5.4in\relax{\includegraphics{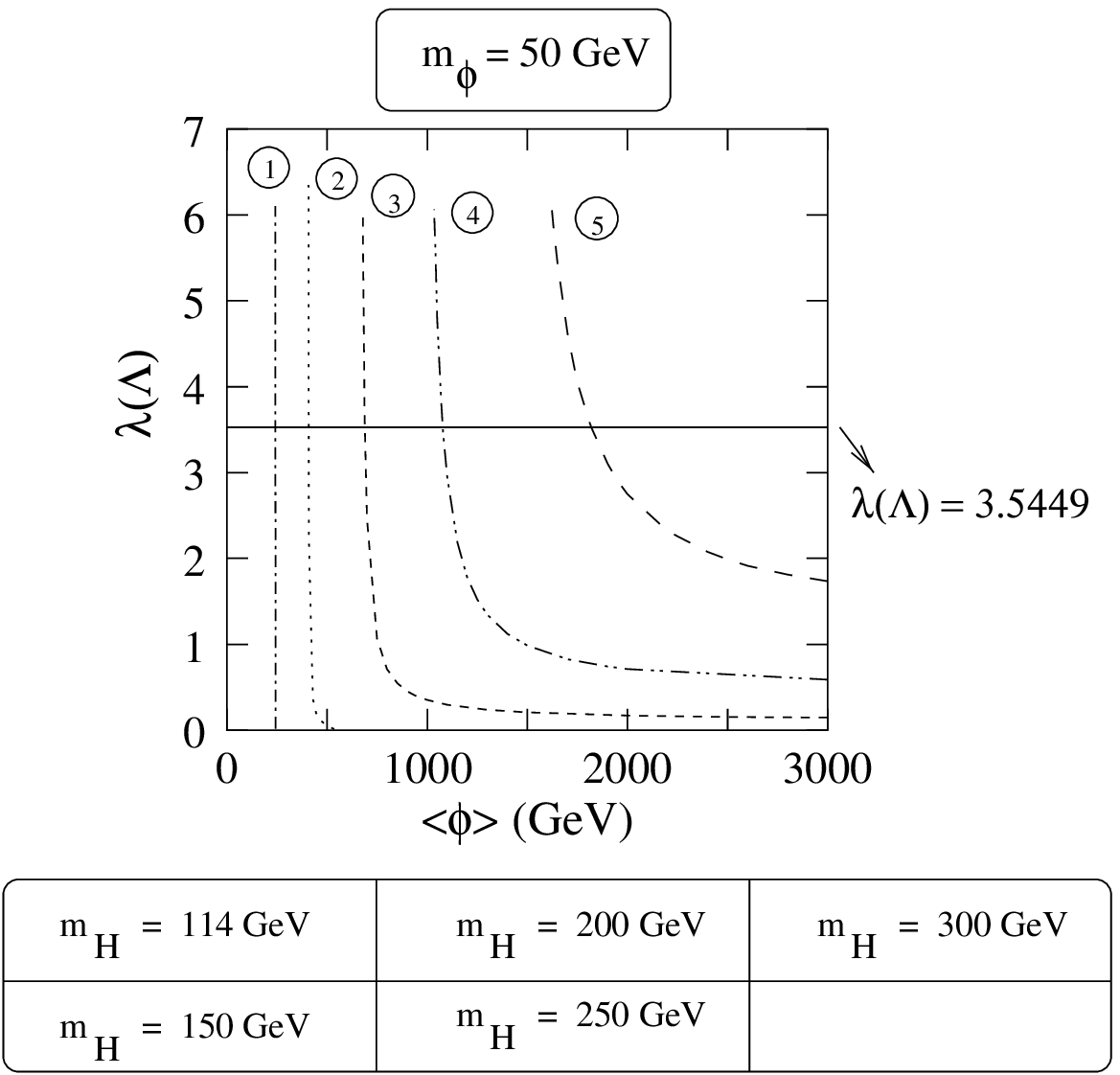}}
\end{center}  
\end{figure}
\vspace*{-0.2in}
\noindent { Figure. 4a}.
{\it {Showing the variation of $g_t(\L)$ against $\vphi$ for different
initial values of $m_h(m_z)$and $m_\phi = 50~ GeV$. }}

\newpage
\begin{figure}[htb]
\begin{center}
\vspace*{4.5in}
      \relax\noindent\hskip -5.4in\relax{\includegraphics{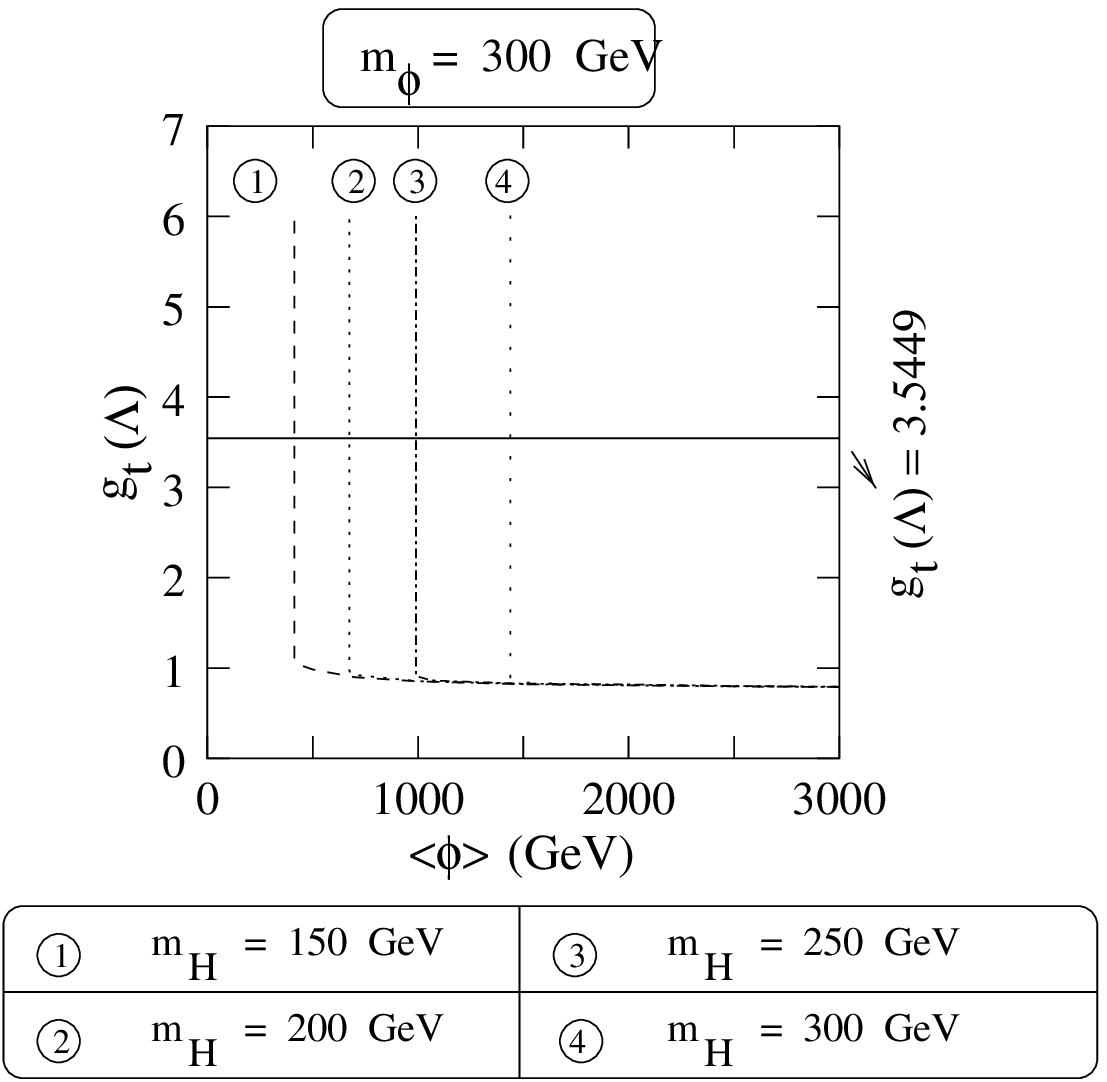}}
\end{center}
\end{figure}
\vspace*{-0.2in}
\noindent { Figure. 4b}.
{\it {The variation of $g_t(\L)$ against $\vphi$ for different 
initial values of $m_h(m_z)$ and $m_\phi = 300~ GeV$. }}

\newpage


\begin{figure}[htb]
\begin{center}
\vspace*{4.0in}
      \relax\noindent\hskip -5.4in\relax{\includegraphics{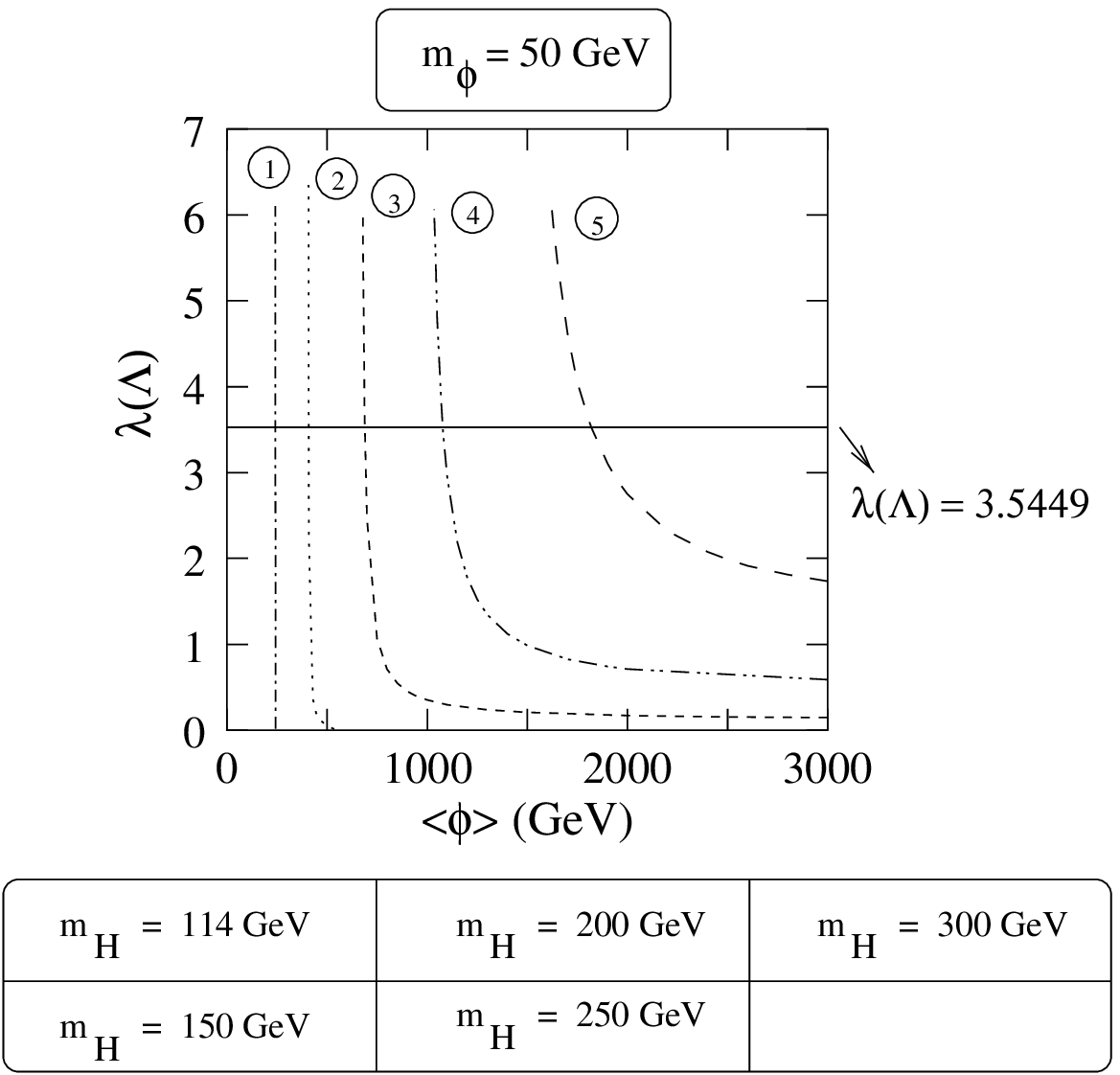}}
\end{center}
\end{figure}
\vspace*{-0.2in}
\noindent { Figure. 5a}.
{\it {Showing the variation of $\l(\L)$ against $\vphi$ for different 
initial values of $m_h(m_z)$ and $m_\phi = 50~ GeV$. }}

\newpage

\begin{figure}[htb]
\begin{center}
\vspace*{4.0in}
      \relax\noindent\hskip -5.4in\relax{\includegraphics{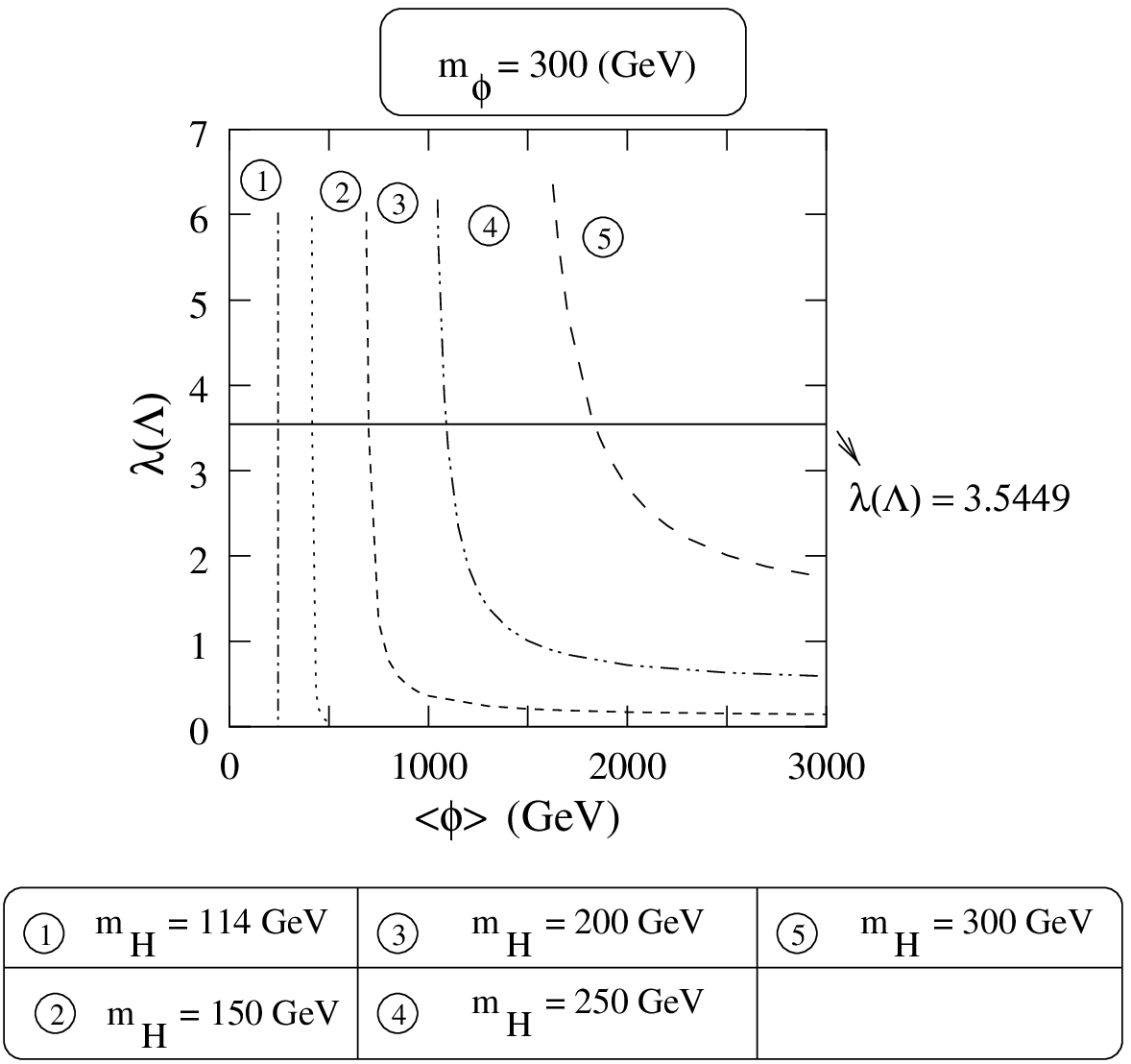}}
\end{center}
\end{figure}
\vspace*{-0.2in}
\noindent { Figure. 5b}.
{\it {The variation of $\l(\L)$ against $\vphi$ for different 
initial values of $m_h(m_z)$ and $m_\phi = 300~ GeV$. }}

\newpage

\vspace*{-1.0in}

\begin{figure}[htb]
\begin{center} 
\vspace*{3.7in}
      \relax\noindent\hskip -4.4in\relax{\includegraphics{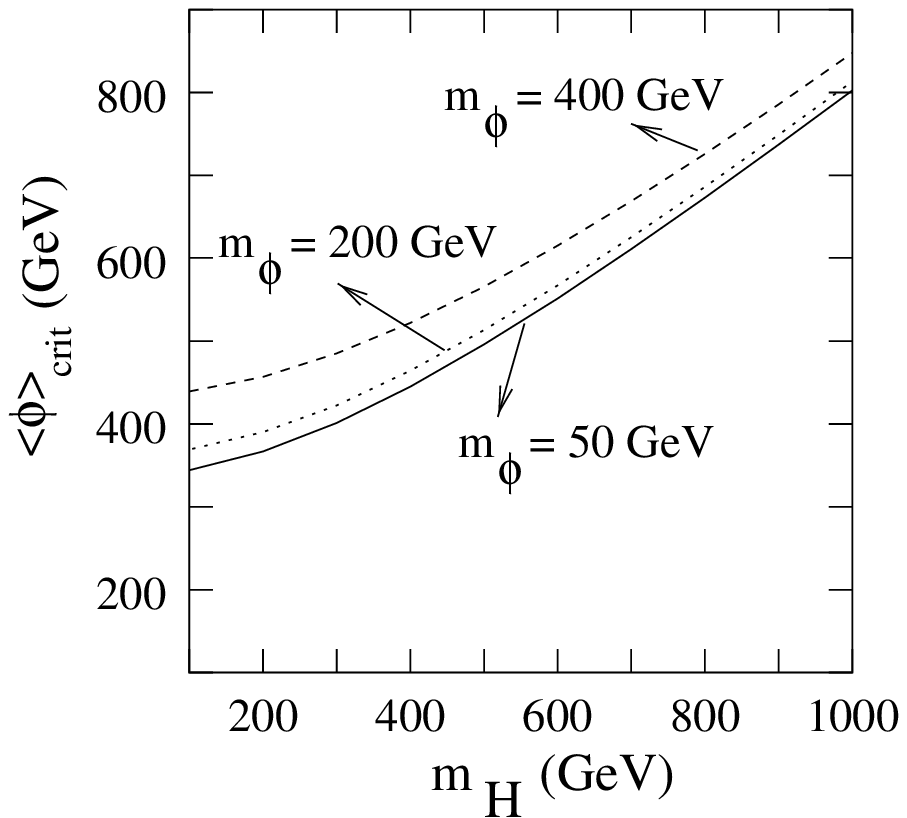}}
\end{center}
\end{figure}
\vspace*{-0.2in}
\noindent { Figure. 6}.
{\it {Showing the variation of $\vphi_{crit}$ against $m_H(m_z)$. }}

\end{document}